\documentclass[runningheads]{llncs}

\usepackage{amssymb}
\usepackage{url}
\usepackage{cite}

\urldef{\mailsa}\path|{franziska.hannss, rainer.groh}@tu-dresden.de| 
\usepackage{graphicx}
\begin{document}

\mainmatter  
\title{What Color is this? Explaining Art Restoration Research Methods using Interactive Museum Installations\let\thefootnote\relax\footnotetext{This paper has been peer–reviewed and accepted to the \textit{Workshop on Visual Interface Design Methods (VIDEM 2020)}, held in conjunction with the International Conference on Advanced Visual Interfaces (AVI 2020). Ischia, Italy. Workshop organizers: Mandy Keck, Dietrich Kammer, Alfredo Ferreira, Andrea Giachetti, Rainer Groh, http://videm.mediadesign-tud.de/}}
\titlerunning{What Color is this? Explaining Art Restoration Research Methods using Interactive Museum Installations}
%
\author{Franziska Hannß\inst{1} \and
Esther Lapczyna\inst{2} \and
Mathias Müller\inst{3} \and
Rainer Groh\inst{1}}
\authorrunning{F. Hannß et al.}
%
\institute{Chair of Media Design, Technical Universtity of Dresden, Dresden, Germany 
\mailsa\\ \and
GTV – Gesellschaft für Technische Visualistik mbH, Dresden, Germany
\email{esther.lapczyna@visualistik.de}\\ \and
University of Applied Sciences Dresden, Dresden, Germany\\
\email{mathias.mueller@htw-dresden.de}}
\toctitle{Lecture Notes in Computer Science}
\tocauthor{Authors' Instructions}
\maketitle              
\begin{abstract}
This case study describes an approach to designing interactive museum installations as a student project with the aim of presenting the research results of the restoration process of paintings to a wide range of visitors. During one and a half years, the Chair of Media Design created five interactive media stations in two lectures to enrich the special exhibition "Veronese: The Cuccina Cycle. The Restored Masterpiece". The project was realised in close communication with the conservators of the Dresden State Art Collections and the employees of the Science and Archaeometric Laboratory of the Dresden University of Fine Arts. The students had to learn about the foreign content and how to translate it into a media–related environment. With suitable teaching methods, we pushed the students towards a deeper understanding of the matter.

\end{abstract}
\section{Introduction}
“A museum [...] acquires, conserves, researches, communicates and exhibits the tangible and intangible heritage of humanity and its environment for the purposes of education, study and enjoyment” \cite{icomWWW}. Therefore, the visitor has the opportunity to engage and explore not only the exhibit but also the collected restoration and research information about the masterpiece in greater depth \cite{marty2008museum}. Nowadays, an interactive museum installation (IMI) \cite{pannier2016can} plays an important role in the digitization strategy of a museum. The design and development process of IMIs starts with the first generated facts about the painting or object. Thus, a new and independent field of research has been opened up, located at the intersection of museology, applied computer science, and media technology. 

\begin{figure}
\includegraphics[width=\textwidth]{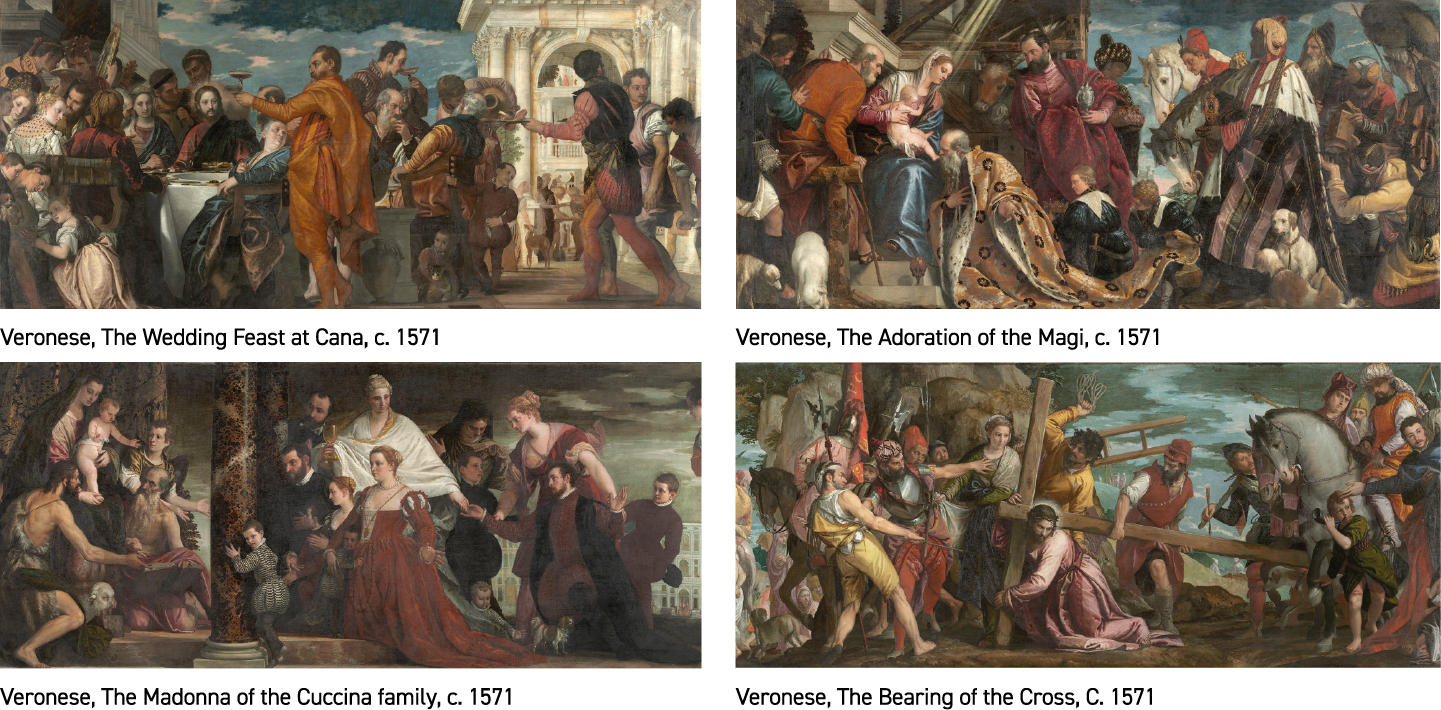}
\caption{Paintings of the exhibition “Veronese: The Cuccina Cycle. The Restored Masterpiece.”} \label{fig1}
\end{figure}

In the following, we describe the design process of IMIs for the special exhibition of the restored paintings of “The Cuccina Cycle” (see
Fig.\ref{fig1}) painted by Paolo Calieri, called Veronese, and how we combined this task with an education project and teaching methods for students at the Chair of Media Design. The exhibition was preceded by a 4–year restoration process. During this process, a wealth of analysis data and new insights of the painting process of Veronese were generated. One focus of the exhibition was to communicate the scientific analysis methods and findings, and to share insights into the restoration process. The presented research material should be accompanied by IMIs to illustrate the complex material and facilitate the understanding as well as present a platform for a conscious dialogue with the visitor. The exhibition took place at the Old Masters Picture Gallery at the Dresden State Art Collections from 9th of March to 3rd of June 2018. \cite{veroneseWWW}

\section{Case Study: Methodical Approach} \label{2 Methodical Approach}
At the beginning of our work the exhibition concept was still work in progress. Due to first restoration results, the content of the exhibition was divided into four focal points, which formed the content of the exhibition and the educational modules:
\begin{itemize}
\item Issue 1 – \textbf{restoration process} – Why was the restoration process necessary? What was part of the restoration process and how did scientific analysis and examination support the restoration decisions?
\item Issue 2 – \textbf{“Madonna of the Cuccina family”} – Painting technique and restoration process through the example of the painting “Madonna of the Cuccina family”
\item Issue 3 – \textbf{New color state after restoration} – What causes the different colour characteristics after the restoration process?
\item Issue 4 – \textbf{Color changing of the pigments} – Information about color change and pigments (smalt, copper, zinc etc.)
\end{itemize}
The museum's work is characterized by a large network of various disciplines. Supporting the knowledge management process in the design of IMI’s is therefore another important aspect of this project. In our experience, a successful knowledge transfer is an essential factor in the development of user–centred teaching and communication approaches.

In order to investigate this process in more detail, we have located the project within the framework of our didactic–methodological research. One of the main focuses of the research and project work of the Chair of Media Design is on the theory and methodology of designing interactive systems. Guided by the scientists, the conceptual and technical implementation was realized as a student semester project. The students were newcomers both in terms of the museum context and the design of user–centred forms of interaction and technologies. The chosen didactically motivated design process uses models and methods from various scientific disciplines and represents a novel combination of modular methodological building blocks. These individual approaches have already been successfully tested in other teaching formats within the scope of the study of media computer science. \cite{brade2013natural,eschrich2013shape,keck2014revisiting, peschke2012depthtouch} Finally, the challenge for the teaching staff was to design a suitable teaching format, which leaves space for research questions and the professional development of IMIs for the final exhibition. 

\subsection{Phase I – The Laboratory}
\begin{figure}
\includegraphics[width=\textwidth]{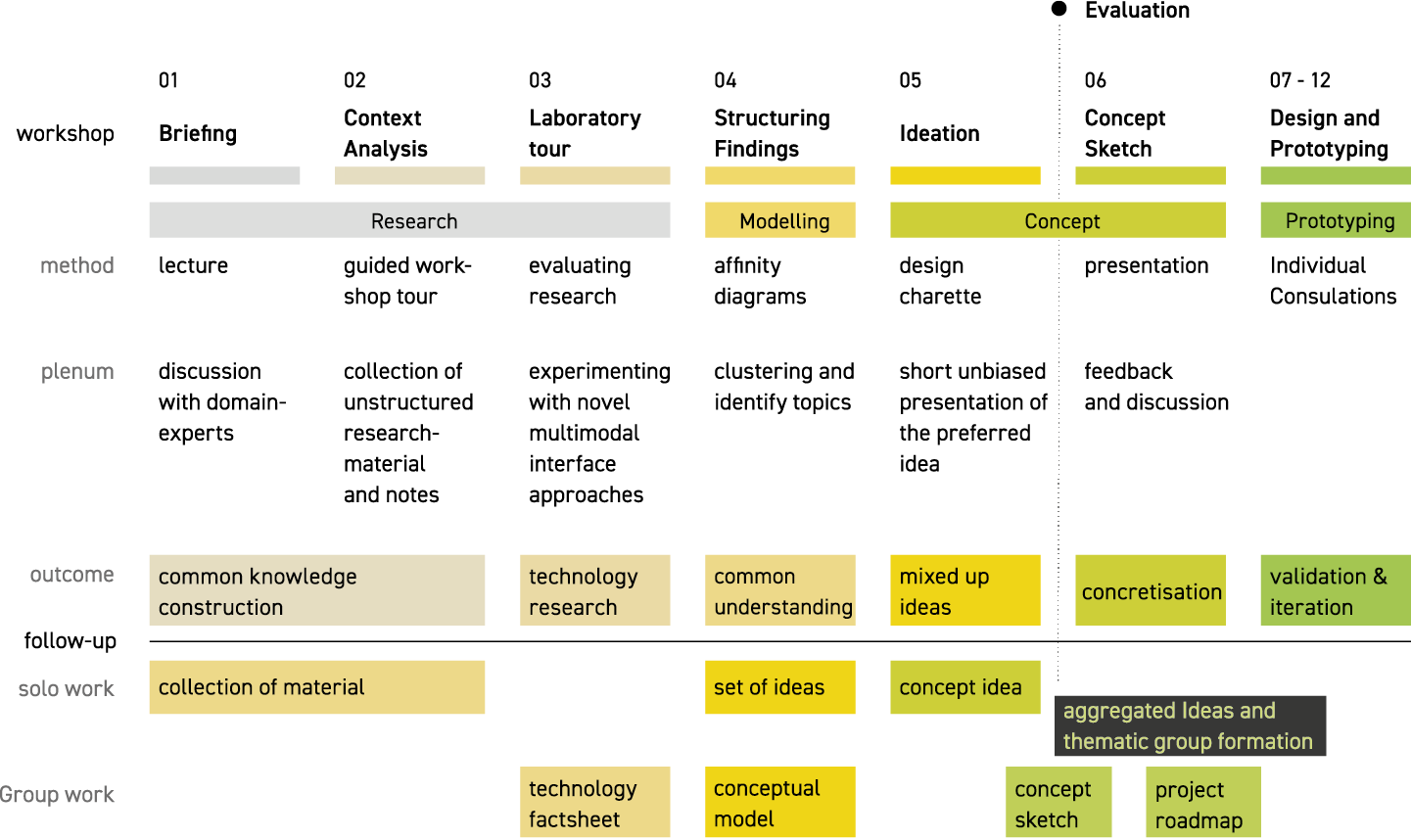}
\caption{Building blocks of the laboratory phase.} \label{fig2}
\end{figure}
The laboratory characterizes the first phase of the project and concluded in a variety of functional prototypes. So far, the research work of the restorers and art historians has provided a rich source of results with focus on the four issues \cite{veroneseWWW}. In consecutive workshop modules, we attempted to strengthen the problem–solving skills of the students within new knowledge domains. A schematic overview of the teaching concept shows the intertwining of the different didactic building blocks (see Fig.\ref{fig2}). With this methodically designed workshops a common knowledge construction and discussion took place. 

The design process is structured into four main units: \textbf{research}, \textbf{modelling}, \textbf{concept}, and \textbf{prototyping}. The \textbf{research unit} was characterised by the exploration of the context and the testing of suitable technologies. Within the scope of a laboratory tour, a valuable exchange of information on the thematic key aspects of the current restoration–process took place. (see fig. \ref{fig2}, workshop 1–3) After structuring all gathered information in the \textbf{modelling unit} the students started to produce a large amount of intentionally imperfect sketches and notes. Using the Design–Charrette–Method \cite{hanington2012universal} at the beginning of the \textbf{concept unit}, the first thoughts of each student went through 12 iterations and at the end of the ideation workshop (see Fig.\ref{fig2}, workshop 5) they formed one final approach. Out of a set of 34 approaches the preferred combination of the ideas and variation was selected based on a quality control of the teaching staff. The student groups were purely formed based on the conceptual proximity of the ideas, which avoided the formation due to friendship. During the \textbf{prototyping unit} the first conceptual prototypes as possible solutions for IMIs were created and presented to the conservators and art historians involved in the exhibition.

\subsection{Phase II – The Consolidation}
\begin{figure}
\includegraphics[width=\textwidth]{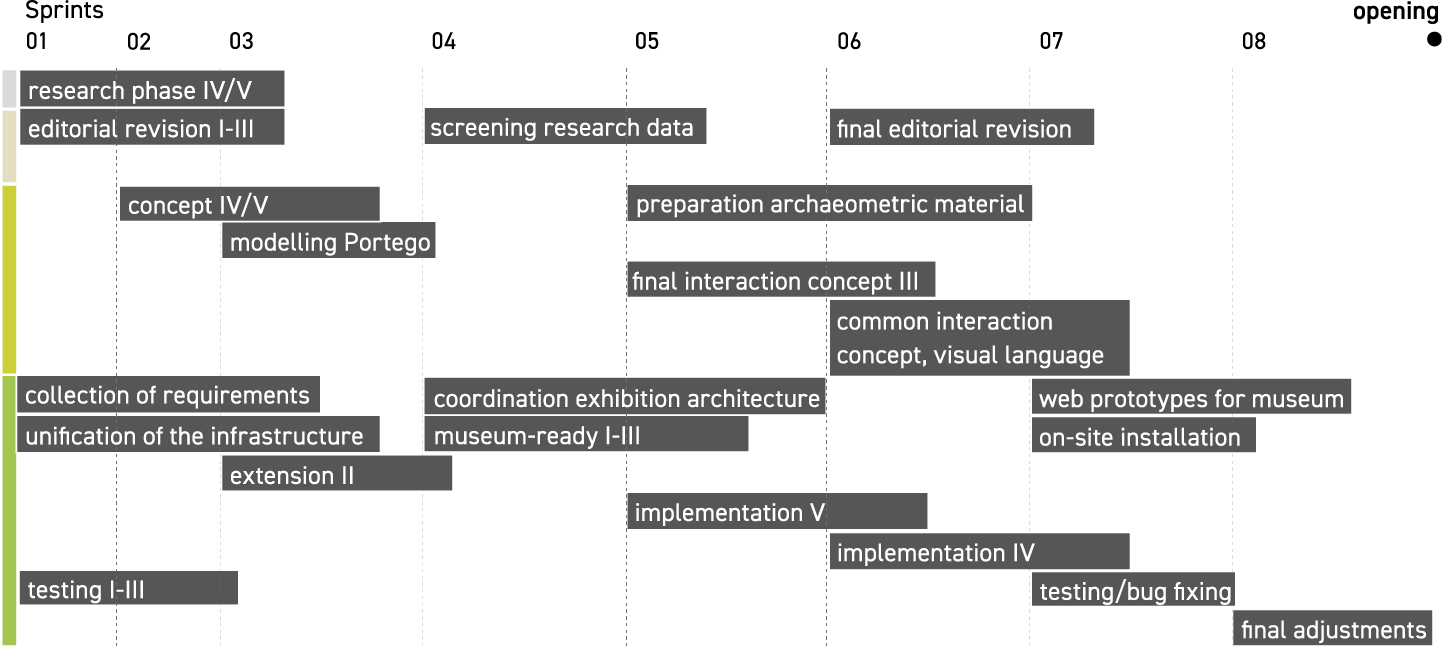}
\caption{Sprint planning of the consolidation phase, categorised in research, editing, design and implementation (top to bottom).} \label{fig3}
\end{figure}
Due to the great variety of functional prototypes, a common knowledge base for the further development of the exhibition design as well as the IMIs was built. Therefore, an extensive catalogue of requirements was created by the media design department for the evaluation of the intermediate results. And with the refinement of the exhibition concept by the museum it was possible to define the final content of the IMIs. Three of the final IMIs consist of a synthesis or extension of the existing prototypes. Two blank spots had been identified, which had to be filled with additional IMI–concepts. 

With Phase II – the consolidation – we transform the teaching method from an prototyping workshop–oriented approach into a practice–oriented development process. Therefore, the development roadmap (see Fig.\ref{fig3}) had to be included in the planing of the exhibition to ensure a target– and museum–oriented implementation of the final concept. The development process is divided in 8 sprints and coordinated by the teaching staff. 

With the consolidation, a new team of four students was formed and supplemented by the teaching staff and colleagues from the museum. In this way all necessary skills were brought together. Two experts from the laboratory phase were supported by two thematic newcomers, who had experience in problem solving strategies and technological expertise with AR and 3D applications. They gained their content–related input from the documented knowledge of the 1st phase. Challenges and Solutions within the development process are described in more detail in section \ref{4. Challenges and Solutions}.

\section[Case Study: Interactive Media Stations]{Case Study: Interactive Media Stations \footnote{A video documentation of the exhibition media stations can be found at: \url{http://www.youtube.com/watch?v=lIZ1N-Ywl5E}}} \label{IMI} 

\begin{figure}
\includegraphics[width=\textwidth]{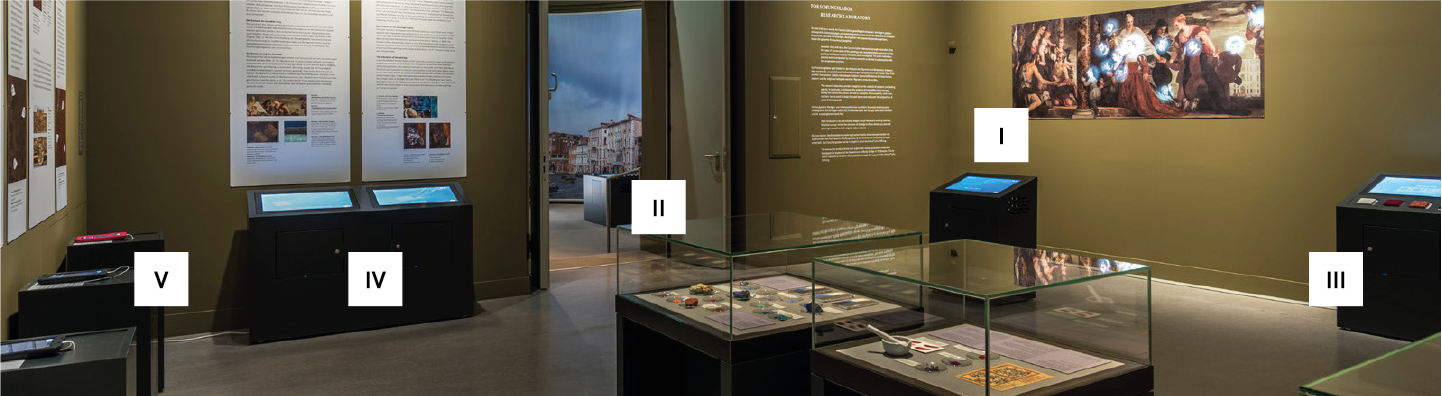}
\caption{Exhibition scenery with the featured IMIs. I – Hidden Relations, II –  The Palazzo Cuccina in Venice, III – Venetian Fabrics and Garments, IV – The Cuccina Cycle by Comparison, V – Exploration of the Original Colorfulness.} \label{fig4}
\end{figure}
\begin{figure}
\includegraphics[width=\textwidth]{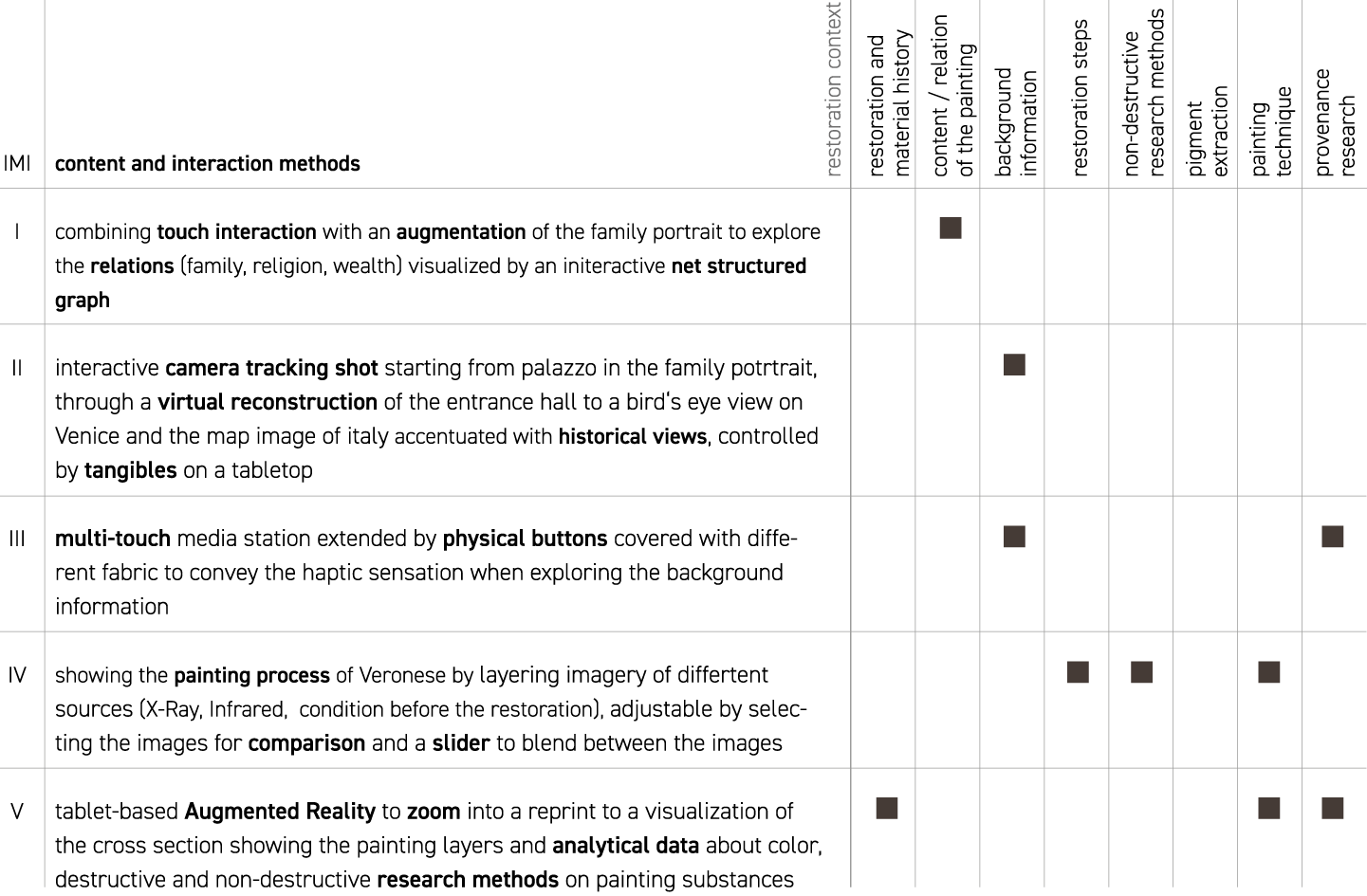}
\caption{Overview of the developed IMIs with the featured interaction methods in relation to the restoration context.} \label{fig5}
\end{figure}

The final exhibition featured five IMIs (see Fig. \ref{fig4}, \ref{fig5}). The media station entitled \textbf{The Palazzo Cuccina in Venice} (II) serves as an introduction to the historical background and is placed in the exhibition’s starting room. It consists of a Microsoft SUR40 multitouch tabletop and hand–crafted tangibles that are used to navigate through the app. 

\textbf{Hidden Relations} (I) consists of a multi–touch tabletop and a projector that projects dynamic spotlights on a reproduction of the painting ``The Madonna of the Cuccina Family``  (see Fig.\ref{fig1}) on the wall in front of the tabletop. These spotlights are controlled by the app, on which the visitor explores the family relationships and thematic links within the portrayed family members and the four virtues.

Another multi–touch media installation – \textbf{Venetian Fabrics and Garments} (III) – is extended by illuminated physical buttons covered with fabric. In addition to multi–touch, they ensure further access to the content.

During the restoration analysis, each painting was photographed using various photographic techniques (macro before and after restoration, x–ray, infrared, color mappings, canvas seams). \textbf{The Cuccina Cycle by Comparison – Restoration and Research} (IV) allow the visitor to explore all four paintings layer by layer.

Finally, we realized a tablet–based Augmented Reality Application using Unity and Vuforia, called \textbf{Exploration of the Original Colorfulness} (V). The visitor can dive into the different analytical processes and findings. A reprint of the corresponding painting area is used as marker for AR tracking. In addition, a reproduction of the original color can be blended virtually over the reprint.

\section{Challenges and Solutions} \label{4. Challenges and Solutions}
Due to the interdisciplinary structure of the project, several challenges had to be overcome in the areas of organisation, communication, complexity and museum context. 
\subsection{Organization and Communication} \label{4.1 Organization}
The whole project had a very generous time frame that allowed us to pass through an extended in–depth concept phase (see Fig. \ref{fig2}: workshop 5 and 6) and aim for a structured and thorough implementation. The final deployment and adaptation of housings, in–place testing and final adjustments for the exhibition was rather stressful (see Fig. \ref{fig3}: sprint 5–8). 

Communication of results and discussing concepts, coordination of the numerous people and interested parties was time–consuming and often different approaches, requirements and constraints had to be satisfied. We therefore used a broad range of tools to overcome these issues – \textit{Trello} for tracking of tasks, arrangements, responsibilities, and deadlines, \textit{Slack} for easy communication, especially between developers and designers, \textit{NextCloud} as cloud storage to share preliminary results and access source materials, \textit{GitLab} as code repository, collaborational development tool and for change tracking, and \textit{Toggl} for time–tracking.

\subsection{Complexity}
The \textbf{thematic complexity} was especially challenging for developers and designers who had to interpret the materials, methods and findings of researchers and restorators. This resulted in an intense research phase (see Fig. \ref{fig2}: workshop 1–3). Furthermore, the available material was very heterogeneous and originates from different sources, resulting in a time–consuming modelling (see Fig. \ref{fig2}: workshop 4) and preparation process (see Fig. \ref{fig3}: sprint 5). As the material was created for scientific documentation purposes with low resolution and different formats, it targeted experts, containing only incomplete, exemplary imagery. Therefore, we put a high effort on selecting and editing the source material for use in the application. 

Subsequently, the \textbf{technological complexity} reflects the thematic complexity posing another demanding challenge as standard consumer solutions had to be adapted to the museum context and the associated technical conditions. Especially projector adjustments and synchronization between different IMI components were challenging (I). On the software side, this resulted in a large number of software frameworks and content–creation tools: Unity3D and Vuforia were used for mobile Augmented Reality (V), Arduino and node.js in conjunction with WebSockets for communication (III) and modern web technologies for visualization and interaction. Further–more, specific adjustments were necessary for Microsoft SUR 40: we used a TUIO–Wrapper to make the Tangibles work together with web technologies (II). The heterogeneous hardware and software was especially challenging for configuring automatic startup, shutdown and restart in case of crashes. The used SiteKiosk needed some sophisticated modifications to support all of the different technologies (I–V).

Finally, the \textbf{design complexity} posed another challenge, especially in terms of maintaining a common look–and–feel and intuitive, consistent interaction concepts (see Fig. \ref{fig3}: sprint 6). Furthermore, the interface design needed to adapt to the surrounding architecture and exhibition design.

\subsection{Museum Context}
\sloppy {To target the wide variety of visitors the IMI had to be intuitive and self-explanatory, invite people to interact, provide stable, consistent, instantaneous feedback/feed–forward as neither manuals nor museum employees are available for providing help. To address the specific stability requirements of intense, daily 8–hour–use, we conducted an intense test phase before deployment to detect crashes and trace down memory leaks (see Fig. \ref{fig3}: sprint 1 and 7).}

The sensitive environment also played a role in the selection of the basic technologies - e.g. AR/VR glasses proved to be impractical, IR light sources had to be used responsibly, interactive stations had to be spatially separated from the original paintings. Furthermore, lighting conditions and energy management for mobile devices in a museum context posed additional challenges.

Further requirements were security and anti-theft protection, which were critical for the tablets in particular and led to some customer-specific hardware modifications. Furthermore, aspects had to be considered when placing the IMIs in the exhibition, such as fire protection, escape routes, maintenance access as well as user routes, cable and energy management, and the installation and accommodation of the hardware.

\section{Conclusion and Lessons Learned}
Looking back onto the design process and observations from the final exhibition, we can formulate some lessons learned which can serve as starting point or best practices for similar projects. Designing for heterogeneous audiences requires a careful use of novel technologies – our hybrid concept to extend established  media types with innovative IMIs proved to be a viable approach.

Another aspect was the extensive use of management tools (see section \ref{4.1 Organization}) during the process to tackle organizational issues. The used tools were effective and really supported the team’s work. However, there is still a lot of potential, especially regarding integration of the different tools in a general workflow. We also lacked instruments for the final project phase and project review, e.g. for advertising purposes or to document and archive the results. The two–phase approach worked very well, especially due to the comfortable time frame. It allowed us to first explore broad possibilities, then take time to analyze material and methodology to create different prototypical  results (mock–ups, rapid prototyping). From our experience, students tend to stick with the first solution, but find it difficult to see the added value of alternatives in early design phases. An unbiased approach to the topic is important. Therefore, early criticism should be avoided. Finally, we could fuse the best concepts in the final implementations. For the students, it provided a large space to explore and also the opportunity to play with different technologies, go through an interdisciplinary workflow, experience open–ended, creative freedom in an exciting field of work.

\subsubsection{Acknowledgement.} We like to thanks Oliver Lenz, Erik Lier, Filip Pižl, Johann Schumacher as well as the student team of the ``Forschungsseminar Veronese``, Marlies Giebe and Wolfgang Kreische from the Dresden State Art Collections and Annegret Fuhrmann from the Science and Archaeometric Laboratory of the Dresden University of Fine Arts.

%
%
%
\bibliographystyle{splncs04}
\bibliography{AVI2020}
%
%
%
%
%
\end{document}